\documentclass[12pt]{article}
\input{epsf}

\def\hm{Higgs mechanism\hspace {.05in}}
\def\lagsb{\mbox{${\cal L}_{\rm SB}\hbox{ }$}}
\def\ra{\rightarrow}

\def\journal{\topmargin 0.0in   \oddsidemargin 0in
        \headheight 0pt \headsep 0pt
        \textwidth 6.5in 
\textheight 9in 
        \marginparwidth 1.5in
        \parindent 2em
        \parskip .5ex plus .1ex         \jot = 1.5ex}
\journal

\begin{document}
\begin{titlepage}

\noindent December 7, 2004      \hfill    LBNL-56734\\

\begin{center}

\vskip .5in

{\large \bf The No-Higgs Signal: Strong WW Scattering at the LHC}
\footnote
{This work is supported in part by the Director, Office of Science, Office
of High Energy and Nuclear Physics, Division of High Energy Physics, of the
U.S. Department of Energy under Contract DE-AC03-76SF00098}

\vskip .2in
Presented at {\em Physics at LHC,} July 13 - 17, 2004, 
Vienna, Austria.\\ To be published in the proceedings. 

\vskip .5in

Michael S. Chanowitz\footnote{Email: chanowitz@lbl.gov}

\vskip .2in

{\em Theoretical Physics Group\\
     Ernest Orlando Lawrence Berkeley National Laboratory\\
     University of California\\
     Berkeley, California 94720}
\end{center}

\vskip .25in

\begin{abstract}
Strong $WW$ scattering at the LHC is discussed as a manifestation of
electroweak symmetry breaking in the absence of a light Higgs boson.
The general framework of the Higgs mechanism --- with or without a
Higgs {\em boson} --- is reviewed, and unitarity is shown to fix the
scale of strong $WW$ scattering. Strong $WW$ scattering is also shown 
to be a possible outcome of five-dimensional models, which do not 
employ the usual Higgs mechanism at the TeV scale. Precision electroweak 
constraints are briefly discussed. Illustrative LHC signals are reviewed 
for models with QCD-like dynamics, stressing the complementarity 
of the $W^{\pm}Z$ and like-charge $W^+W^+ + W^-W^-$ channels.

\end{abstract}

\end{titlepage}

\renewcommand{\thepage}{\roman{page}}
\setcounter{page}{2}
\mbox{ }

\vskip 1in

\begin{center}
{\bf Disclaimer}
\end{center}

\vskip .2in

\begin{scriptsize}
\begin{quotation}
This document was prepared as an account of work sponsored by the United
States Government. While this document is believed to contain correct
information, neither the United States Government nor any agency
thereof, nor The Regents of the University of California, nor any of their
employees, makes any warranty, express or implied, or assumes any legal
liability or responsibility for the accuracy, completeness, or usefulness
of any information, apparatus, product, or process disclosed, or represents
that its use would not infringe privately owned rights.  Reference herein
to any specific commercial products process, or service by its trade name,
trademark, manufacturer, or otherwise, does not necessarily constitute or
imply its endorsement, recommendation, or favoring by the United States
Government or any agency thereof, or The Regents of the University of
California.  The views and opinions of authors expressed herein do not
necessarily state or reflect those of the United States Government or any
agency thereof, or The Regents of the University of California.
\end{quotation}
\end{scriptsize}

\vskip 2in

\begin{center}
\begin{small}
{\it Lawrence Berkeley National Laboratory is an equal opportunity employer.}
\end{small}
\end{center}

\newpage

\renewcommand{\thepage}{\arabic{page}}
\setcounter{page}{1}

\noindent {\bf 1. Introduction}

Fifty years of high energy physics have led us to a fundamental
question --- {\em What breaks electroweak symmetry?} --- that differs
from other fundamental questions in one respect: {\em we know how to
find the answer!} The way to the answer is to build and run the
LHC. The ability to observe strong $WW$ scattering is an essential
part of this prescription.  If we do observe it, we learn that
elecroweak symmetry breaking is accomplished by strongly coupled
quanta above 1 TeV. If we do not observe it (and know we could have it
if it were present) then we can conclude that elecroweak symmetry
breaking is due to weakly coupled quanta below 1 TeV, which will be
Higgs bosons if the Higgs mechanism is valid.

If we do not initially see light Higgs bosons or strong $WW$
scattering, the confirmed absence of  strong $WW$ scattering would be a
signal to look harder below 1 TeV rather than to expand the search to
scales much greater than 1 TeV. This is a key difference with respect
to many other searches for new physics, where failure to find the
signal at a given energy typically sends us off to search at
still higher energies. The ability to observe strong $WW$
scattering confers a ``no-lose'' capability to determine the mass
scale of electroweak symmetry breaking physics.

Even if a light Higgs boson is discovered, it
will still be important to measure the $WW$ scattering cross section
in the TeV region.  If symmetry breaking is due to a light Higgs
boson, a central prediction of the Higgs mechanism is that strong $WW$
scattering does {\em not} occur.  As discussed below, strong $WW$
scattering is first-cousin to the famous ``bad high energy behavior''
of massive vector boson scattering, which it is a principal mission of
the Higgs mechanism to remove.  If electroweak symmetry breaking is
driven by a strong interaction, the cross section for scattering of
longitudinally polarized $W$ bosons grows toward the unitarity upper
limit, while for symmetry breaking by a weak force it cuts off while
it is still small, well below where unitarity would be saturated.  In
considering the experimental signals at the
\begin{figure}
\centerline{\epsfxsize= 0.7\textwidth \epsfbox {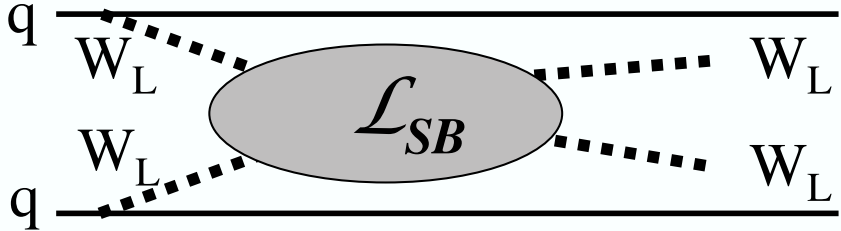}}
\caption{$W_LW_L$ fusion.}
\end{figure}
\noindent 
LHC we should consider
both the capability to observe strong $WW$ scattering if it is present
and to exclude it if it is not.

The basic idea is that we have already discovered three
quanta from the Higgs sector: the longitudinal spin modes of the
$W^{\pm}$ and $Z$ bosons, which in the Higgs mechanism are essentially
Higgs sector quanta.  By measuring the scattering of the longitudinal
modes $W_{L}W_{L} \rightarrow W_{L}W_{L}$ (where $L$ denotes
longitudinal) we are probing Higgs sector interactions,\cite{mcmkg} a
statement made precise by the `equivalence theorem.'\cite {et1,et2} At
the LHC we look for $W_{L}W_{L}$ pairs produced by $WW$ fusion, shown
in figure 1: the $W_{L}$ bosons emitted by the colliding quarks are
off-shell and must rescatter to emerge on-shell in the final state.
Rescattering of electroweak strength is assured by the
interactions of the electroweak Lagrangian ${\cal L}_{EW}$; these 
$WW$ pairs are predominantly transversely polarized. If 
the symmetry breaking sector ${\cal L}_{SB}$ is strongly interacting, 
then the yield is significantly enhanced by the production of 
longitudinally polarized $W_{L}W_{L}$ pairs, indicated in figure 1.

\noindent {\bf 2. The Higgs {\em mechanism}}

The ingredients of the \hm are a gauge sector and a symmetry 
breaking sector,\cite{zuoz}
$$
{\cal L} = {\cal L}_{\hbox{\small{gauge}}} + \lagsb.
\eqno(2.1)
$$
${\cal L}_{\hbox{\small{gauge}}}$ has an unbroken {\em local} $SU(2)_L \times
U(1)_Y$ symmetry, with massless, transversely polarized gauge
bosons $W^{\pm}$, $Z$, and $\gamma$. \lagsb is the symmetry breaking
Lagrangian that describes the dynamics of the symmetry breaking force
and the associated quanta. The Higgs mechanism requires \lagsb to have 
a {\em global} symmetry group $G$ that breaks spontaneously to a subgroup $H$,
$G \to H$. We do not know $G$ or $H$ but we do know that 
$G \supset SU(2)_L \times U(1)_Y$ and $H \supset  U(1)_{EM}$, which 
ensures that the resulting Goldstone bosons include 
three, $w^{\pm}$ and $z$, that couple to the three gauge currents 
corresponding to the three spontanteously broken symmetries of 
$SU(2)_L \times U(1)_Y$. The Higgs mechanism then ensures that 
$w^{\pm}$ and $z$ become the longitudinal modes of the gauge 
bosons $W^{\pm}$and $Z$ which acquire masses, {\em whether there is 
a Higgs boson or not}.

The equivalence theorem\cite{et1}, valid to all orders in 
gauge and symmetry breaking \\
interactions,\cite{et2}  codifies the 
fact that the longitudinal gauge boson modes, $W^{\pm}_{L},Z_{L}$,  
behave as quanta from \lagsb at high energy, $E \gg m_W$, 
$$
{\cal{M}}(W_L(p_1), W_L(P_2), \ldots ) = {\cal{M}} (w(p_1), w(p_2),\ldots )_R
+O\left({m_W\over E_i}\right),\eqno(2.2)
$$
where the subscript $R$ denotes a covariant renormalizable gauge choice 
such as Landau gauge. Equation 2.2 underlies the statement that 
$WW$ fusion probes \lagsb as indicated in figure 1. 

Goldstone boson scattering obeys low energy theorems (LET's), as  shown 
by Weinberg for $\pi\pi$ scattering. The  LET's for $w,z$ can be derived 
without knowing $G$ and $H$; for instance\cite{mcmghg}
$$
{\cal{M}}(w^+w^- \to zz) = {1\over \rho}{s\over v^2},\eqno(2.3)
$$
where $v$ and $\rho$ are the usual vev and rho parameter. 
Equation (2.3) is valid 
at low energy, $s \ll\ M^2_{\rm SB}$,  
where $M_{\rm SB}$ is the typical mass scale of $\lagsb$. Combining the 
LET (2.3) and the ET (2.2) we obtain the gauge boson LET,
$$
{\cal{M}}(W_L^+W_L^-\to Z_LZ_L)={1\over\rho}{s\over v^2}\eqno(2.4)
$$
valid in the intermediate energy 
domain $m^2_W \ll s \ll M^2_{\rm SB}$.\cite{mcmghg}

To understand strong $WW$ scattering it is instructive to consider 
a U-gauge derivation of the LET which makes no reference to the 
underlying Goldstone bosons or the ET.\cite{mcmghg} In leading order 
the U-gauge amplitudes involving only gauge sector quanta 
exhibit the ``bad'' high energy behavior that would
make massive vector boson theories nonrenormalizable.  
E.g., with just the gauge sector Feynman diagrams we have at high energy 
$$
{\cal{M}}(W_L^+W_L^-\to Z_LZ_L)_{\rm gauge\ sector}=
   {g^2s\over 4\rho m_W^2} +O(g^2s^0) \simeq {s\over \rho v^2}, \eqno(2.5)
$$ 
which would eventually violate unitarity and render the theory 
nonrenormalizable. 
This ``bad'' behavior is cancelled at the scale $M_{\rm SB}$ by
exchange of quanta from ${\cal L}_{\rm SB}$, 
but at low energy, $s \ll M^2_{\rm
SB}$, it can be shown that \lagsb decouples to all orders. Therefore we
again obtain the LET (2.4) for $m^2_W \ll s \ll M^2_{\rm SB}$.  
We see that {\em the LET is precisely the low energy tail of 
the ``bad'' UV behavior.}

There are two important conclusions from this discussion:
\begin{itemize}
\item In the Higgs mechanism, $W_L W_L$ scattering exhibits the 
Goldstone boson dynamics of \lagsb.
\item Even if the Higgs mechanism does not occur we see from the U-gauge 
derivation that the $W_L W_L$ LET is still valid if the physics \lagsb 
that cuts off the amplitude decouples at low energy.
\end{itemize}
\begin{figure}
\begin{center}
{\epsfxsize= 0.7\textwidth \epsfbox {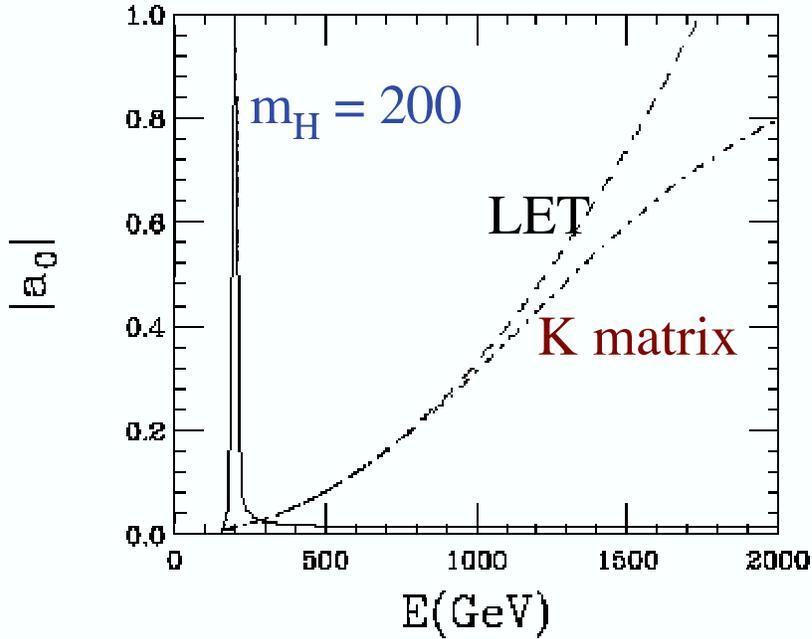}}
\end{center}
\caption{$|a_0(W_L^+ W_L^-\to Z_LZ_L)|$ for $m_H= 200$ GeV
and two strong scattering models.}
\end{figure}

\noindent {\bf 3. Unitarity and the scale of Strong $WW$ scattering}

Unitarity implies a rigorous upper bound on the energy at which quanta
from \lagsb must cut off the growth of $W_L W_L$ scattering.  For
example, setting $\rho = 1$ (assuming \lagsb has a custodial isospin
symmetry $I$), the LET for the $I=J=0$ partial wave is
$$
|a_{00}(W_L W_L)| = {s\over 16\pi v^2}. \eqno(3.1)
$$
Below four-body thresholds unitarity requires $|a_{00}| \leq 1$ and 
${\rm Re}\ a_{00} \leq 1/2$, which would be violated at 1.8 and 1.2 TeV 
respectively. These values imply that $M_{\rm SB}$ cannot be too 
much greater than 1 TeV, say $M_{\rm SB} \leq O(2)$ TeV. 

The existence of strong $WW$ scattering then depends on the the scale 
of  $M_{\rm SB}$. If \lagsb is weakly coupled, then $M_{\rm SB} \ll 1$ TeV
and $WW$ scattering never becomes strong. For instance, consider the 
SM with a light Higgs boson, $M_{\rm SB}=m_H \ll 1$ TeV. 
Then ${\cal M} \simeq  s/v^2\ (1-s/(s-m_H^2))$ where the second term, 
from $H$ exchange, cancels the first term (2.5) from gauge sector 
interactions. At high energy the sum is $\simeq m_H^2/v^2 = 2\lambda_H$ 
and $a_{00} \simeq (m_H/1.8 {\rm TeV})^2$. On the other hand if 
$M_{\rm SB}$ is greater than 1 TeV  then the partial wave amplitudes 
rise toward their unitarity limits, $a_{00} \simeq O(1)$, for 
$E>1$ TeV, so that the scattering is strong. This is illustrated in 
figure 2, which shows $|a_0(W_L^+ W_L^-\to Z_LZ_L)|$ 
for two strong scattering models (LET and K-matrix) compared to 
the SM with $m_H= 200$ GeV.

\begin{figure}
\begin{center}
\centerline {\epsfxsize= 0.8\textwidth \epsfbox {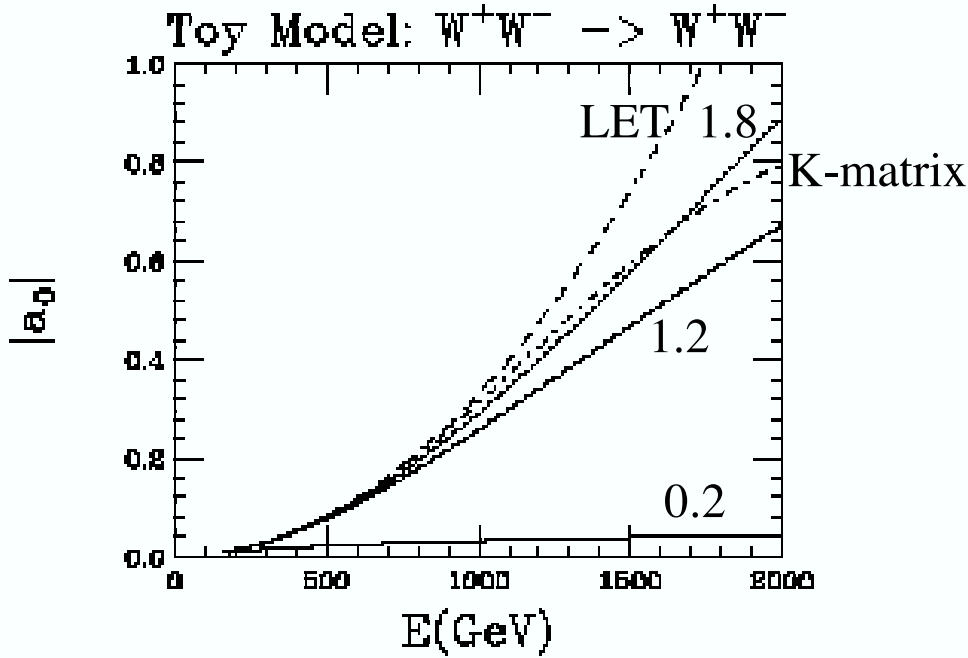}}
\end{center}
\caption{$|a_0(W_L^+ W_L^-\to W_L^+ W_L^-L)|$ for 5-d model with 
$m_1=0.2, 1.2, 1.8$ TeV.}
\end{figure}

\noindent {\bf 4. No Higgs mechanism? --- EWSB in 5 dimensions}

The Higgs mechanism has been an article of faith in high energy 
physics for about 30 years but it has not been tested experimentally. We 
will test it at the LHC. The experimental success of 
the SM implies that ${\cal L}_{SU(2)\times U(1)}$ is a good effective 
theory below the scale of new physics, {\em even if the Higgs mechanism 
does not occur in  nature.} From the U-gauge derivation  we know that 
the $WW$ low energy theorems are valid in this  case, and 
unitarity would require SOMETHING to cut off $a_0(W_L W_L)$. If 
$\Lambda_{\rm SOMETHING} \geq O(1)$ TeV then strong $WW$ scattering 
would occur, just as it would for the Higgs mechanism with 
$M_{\rm SB}>1$ TeV.

An example of ``SOMETHING'' has recently emerged from
extra-dimensional theories. Models in which ${\cal L}_{SU(2)\times
U(1)}$ is broken by boundary conditions on a compact fifth dimension
can postpone the violation of unitarity to tens of TeV. The
cancellation of bad UV behavior at the 1 TeV scale is accomplished by
exchange of the Kaluza-Klein excitations of the gauge
bosons,\cite{cdh} $W_n,Z_n,\gamma_n$, with masses $M_n \simeq n/R$
where $R$ is the size of the compact dimension.

Yang-Mills theory in five dimensions is nonrenormalizable and can only
be an effective theory below a cutoff $\Lambda_5$. It is 
possible for $\Lambda_5$ to be an order of magnitude larger 
than the  TeV scale, in which case unitarity in the effective 
four-dimensional theory is preserved up to $\Lambda_5$ by 
cancellations from exchanges of the Kaluza-Klein gauge 
bosons.\cite{cdh} Above $\Lambda_5$ the physics underlying the five
dimensional theory (strings?) would emerge.

At the LHC $W_LW_L$ scattering could be weak or strong depending on
the mass of the KK bosons. For $M_1 \ll 1$ TeV it would be weak while
for $M_1 \simeq $ O(TeV) it would be strong. To illustrate this I have
considered $W_L^+ W_L^-\to W_L^+ W_L^-$ scattering in a toy model
(actually the Georgi-Glashow model) considered by Csaki {\em et
al.}\cite{csakietal} with symmetry breaking $SU(2) \to U(1)$. The bad
high energy behavior of the gauge sector amplitudes is cancelled in
this case by the $s$ and $t$ channel exchanges of $\gamma_1$, the
first Kaluza-Klein excitation of the photon. By varying the radius $R$
of the compactified fifth dimension we can specify the mass $M_1$ of
$\gamma_1$.  Figure 3 compares the results for $M_1 = 0.2$, 1.2, and
1.8 TeV with the two strong scattering models shown in figure 2. For
$M_1 = 0.2$ TeV scattering is weak, as for a light Higgs boson, but
for $M_1 > 1$ TeV there is strong $WW$ scattering as the partial waves
grow toward O(1) above 1 TeV.

The strongly coupled versions of these models are preferred
experimentally because they are better equiped to hide both from
precision electroweak constraints\cite{bn} and from direct KK
searches. These strongly coupled models resemble technicolor, but the
flexibility of the extra-d scenarios give them better prospects to
incorporate fermion masses without large flavor changing neutral
currents.\cite{gb}

Professor Higgs may however have the last laugh: the leading candidate
for this class of models, $AdS_5$, has a dual $CFT_4$ description, in
which symmetry breaking is driven by a strong gauge force, i.e.,
technicolor in a $CFT_4$ setting.\cite{csakietal2} In this case
dynamical symmetry breaking {\em alla} the Higgs mechanism emerges as
the fundamental four dimensional description. And even in the 5-d
theory a version of the Higgs mechanism may be at play: the fifth
components of the gauge fields act like the Goldstone bosons that
become longitudinal gauge boson modes\cite{cdh} and, extrapolating
from a study of brane-induced SUSY breaking, it appears that Wilson
integrals looping over the fifth dimension may provide the symmetry
breaking condensates.\cite{bagger}

\noindent {\bf 5. Precision electroweak constraints}

The SM fit of the precision electroweak data favors a light Higgs
boson with $m_H < 240$ GeV at 95\% CL, implying that there would not
be strong $WW$ scattering at the LHC. But going beyond the Standard
Model, new physics could raise the scale of $m_H$ arbitrarily, even
into the realm of dynamical symmetry breaking, as discussed below.
Furthermore, the longstanding $3\sigma$ discrepancy between $A_{LR}$
and $A_{FB}^b$, the two most important asymmetry measurements in the
SM fit of $m_H$, raises questions about the reliability of the SM
determination of $m_H$.\cite{mscafbb} {\em LEP cannot definitively 
determine the scale of electroweak symmetry breaking. LHC can.}

The discrepancy between $A_{LR}$ and $A_{FB}^b$ could be a genuine
manifestation of new physics. If it is, the SM fit is moot and we
cannot predict $m_H$ from the precision data until the new physics is
known.  If on the other hand the discrepancy is the result of
underestimated systematic uncertainty in the $A_{FB}^b$ measurement
(and the two lower precision hadronic asymmetry measurements,
$A_{FB}^c$ and $Q_{FB}$ --- see \cite{mscafbb}), then the SM
prediction for the Higgs boson mass is very low, $m_H = 58$ GeV, and
\begin{figure}
\begin{center}
\centerline{\epsfxsize= 0.7\textwidth \epsfbox {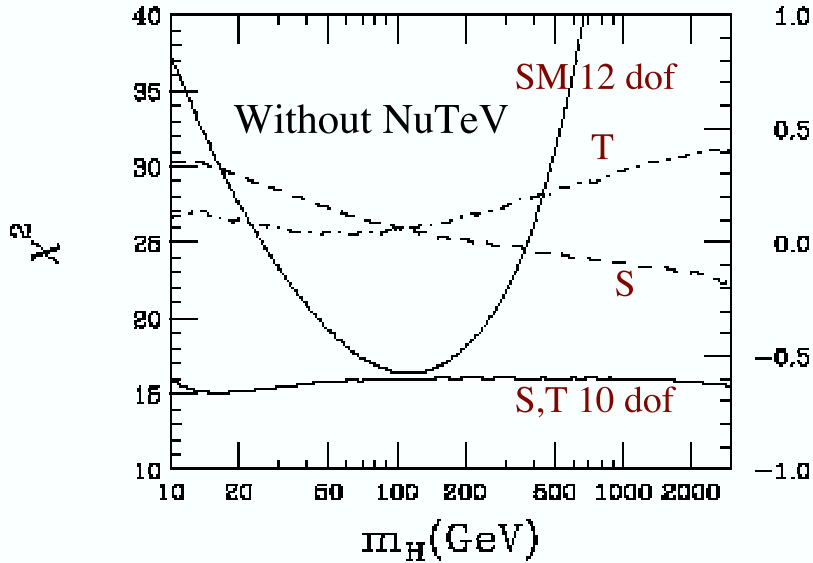}}
\end{center}
\caption{SM and new physics fits to precision data. $S,T$ are read to 
the right axis.}
\end{figure}
is inconsistent at 90\% CL with the LEP II lower limit, $m_H>114$
GeV. In 
this case new physics would be required to raise the predicted
value of $m_H$ into the experimentally allowed region, and again we
could not predict $m_H$ from the precision data until the new physics
were known. To sustain the SM prediction and upper limit for $m_H$, the 
$3\sigma$ $A_{LR}$ - $A_{FB}^b$ discrepancy cannot be new physics or a 
systematic effect but must be a statistical fluctuation. 
This is surely possible, 
but so are new physics or underestimated systematic uncertainty. 
In the latter two cases the precision data does not give us any 
information about the scale of EWSB and strong $WW$ scattering 
remains a possibility. 

Even if the discrepancy between $A_{LR}$ and $A_{FB}^b$ is a
statistical fluctuation so that the SM fit is acceptable at face
value, it is possible that the data could also be explained by
new physics models with {\em any} value of $m_H$. This can be seen
explicitly in models in which the new physics is ``oblique,'' i.e.,
contributes dominantly via corrections to the gauge boson two-point
functions. Figure 4, from \cite{mscafbb}, shows both the SM fit to
the precision data and a new physics fit using the oblique parameters
$S,T$\cite{peskin-takeuchi}. The $\chi^2$ distribution for the $S,T$ fit 
is almost flat, with no significant preference for any value of $m_H$
between 10 and 3000 GeV. Using the current Summer 2004 data, the
confidence level for the oblique fit is 0.17, not very different from
the confidence level for the SM fit, which is 0.23 at the 
$\chi^2$ minimum. (The NuTeV data is not included in these fits; 
they correspond to ``fit C'' from \cite{mscafbb}.)

In figure 4 for each $m_H$ the values of $S,T$ at the $\chi^2$
minimum are shown, plotted against the right vertical axis. To reach
the domain of dynamical symmetry breaking and strong $WW$ scattering
we require negative $S$ and positive $T$. Positive $T$ is readily
available in new physics models --- it corresponds to breaking of
custodial isospin --- but negative $S$ is harder to come by, although
models with $S<0$ do exist.\cite{negativeS} Recently models of the 
type discussed above, with 
electroweak symmetry breaking from boundary conditions on a compact 
fifth dimension, have been formulated with negative $S$ of the 
required magnitude.\cite{csakinegS} However in a broad class of such models 
Chivukula {\em et al.}\cite{chivukulaS} have obtained the 
interesting inequality
$$
S-4{\rm cos}^2\theta_W\ T= {4\over \alpha} {\rm sin}^2\theta_W
        {\rm cos}^2\theta_W\ m_Z^2 \Sigma_n {1\over M_n^2} 
        \geq {1 \over 3}            \eqno(5.1)
$$ 
where the sum is over the Kaluza-Klein excitations of the $Z$ boson
with mass $M_n$. The lower limit follows because $M_1$ cannot be
arbitrarily large in order to unitarize $WW$ scattering, as described
in the preceding section. It then appears that negative $S$ implies
negative $T$ in these models, contrary to what is needed in figure
4. However the values of $S,T$ in these calculations are not
equivalent to the experimental $S,T$, since the latter are normalized
with respect to the SM with a reference value for $m_H$ while there is
no Higgs boson at all in these 5-d models.\cite{csakietal2}
It is an open problem to extract oblique parameters from the 5-d
models that can be compared directly with experiment.

\noindent {\bf 6. Signals at LHC: complementarity in a QCD'ish example}

Strong $WW$ scattering is among the most challenging physics goals 
of the LHC, requiring the full energy and luminosity. Theorist-level 
simulations\cite{mcwk1, mcwk2, baggeretal} indicate that signals 
will be observable at the $\geq 5\sigma$ level with $\geq 150fb^{-1}$ 
data samples. Studies with realistic detector simulations have been 
done by ATLAS\cite{atlas} and CMS\cite{cms}, but it is fair to say 
that simulation studies are still in early days. 

As a generic example I will briefly review the results of a
study\cite{mcwk1, mcwk2} of strong $WW$ scattering signals in
$SU(N_{TC})$ technicolor, using a scaled version of a chiral effective
Lagrangian model of $\pi$ and $\rho$ interactions\cite{weinberg},
which is also the basis of the BESS model\cite{bess} of strong EWSB.
The partial wave amplitudes obtained from the effective Lagrangian are
unitarized by the K-matrix method, which is equivalent to the
conventional Breit-Wigner resonance parameterization with the
constant imaginary part of the denominator, $m \Gamma$, replaced by
$\sqrt{s} \Gamma(\sqrt{s})$.

In \cite{mcwk1} the model was shown to give a surprisingly good
description of the $I=J=1$ and $I=2$, $J=0$ $\pi \pi$ partial
waves. The amplitudes are determined with no free parameters from the
known values of $F_{\pi}$, $m_{\rho}$, and $\Gamma_{\rho}$.  The
results are shown in figure 5.  While the quality of the fit at 1.2
GeV is probably fortuitous, we can take seriously the qualitative
agreement of the model with the data. The $\rho$ meson exchange
obviously enhances the signal in the $a_{11}$ partial wave. Less
obviously $\rho$ exchange {\em suppresses} the $a_{20}$ partial wave,
causing the slope to begin to flatten out at $\simeq 800$ MeV, in
accord with the data shown in figure 5b.  It is then useful to use
the model to explore the consequences of varying the $\rho$ mass and
width. In particular we find that this enhancement/suppression from
$\rho$ exchange implies that LHC signals from the $a_{11}$ and
$a_{20}$ partial waves are {\em complementary}: as $m_{\rho}$ is
decreased the signal at the LHC from $a_{11}$ is enhanced while the
signal from $a_{20}$ is suppressed. Conversely for large $m{\rho}$, 
$a_{20}$ is enhanced while $a_{11}$ is suppressed.
\begin{figure}
\begin{center}
\centerline {\epsfxsize= 0.9 \textwidth \epsfbox {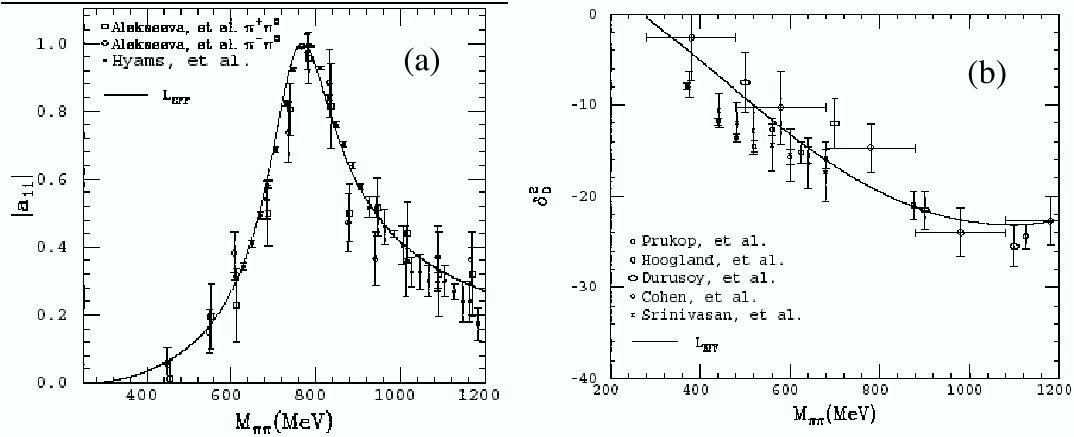}}
\end{center}
\caption{$\pi \pi$ partial waves from chiral Lagrangian.\cite{mcwk1}.}
\end{figure}

The $\rho$ mass and width for $SU(N_{TC})$ are obtained from the
hadronic $m_{\rho},\Gamma_{\rho}$ by the scale factor $v/F_{\pi}\simeq
2700$ modified by factors of $N_{TC}/3$ using the large $N$
scaling laws for $SU(N)$ gauge theories.  For $N_{TC} = 2$
and 4 we have $m_{\rho}= 1.78$ TeV and 2.52 TeV respectively. To
represent the possibility that the resonances of \lagsb may be heavier
than the naively anticipated 1 - 3 TeV region a 4 TeV $\rho$ meson is
also considered, with its width determined from the hadronic $f_{\rho
\pi \pi}$ coupling.

The complementarity of the $a_{11}$ and $a_{20}$ channels is evident
in figure 6. As $m_{\rho}$ increases, the amplitudes approach the
nonresonant K-matrix model amplitude for $|a_{11}|$ from above and
$|a_{20}|$ from below, since chiral invariant $\rho$ exchange enhances
the former and suppresses the latter.  At the LHC the 4 Tev $\rho$
signal is indistinguishable from the signal of the nonresonant
K-matrix model.  The fact that the $\rho$ resonance amplitude
approaches the nonresonant K-matrix amplitude for large $\rho$ mass is
a very general feature, independent of the specific properties of
vector meson exchange.  It explains the sense in which smooth
unitarization models, such as the linear and K-matrix models, are
conservative: they represent the ``fail-safe'' nonresonant scattering
signals that are anticipated if the resonances are unexpectedly heavy.
This is the most general meaning of complementarity.  A more specific
meaning, special to vector meson exchange as constrained by chiral
symmetry, is the inverse relationship of the $I=1$ and $I=2$ channels
discussed here.

The experimental signals and backgrounds for the model were computed
in \cite{mcwk1,mcwk2}. The final states are $W^{\pm}Z$, which includes
the direct channel $\rho^{\pm}$ resonance, and $W^+W^+ + W^-W^-$,
which is pure $I=2$. The irreducible backgrounds are the SM
$\hbox{O}(\alpha_W^2)$ and $\hbox{O}(\alpha_W\alpha_{S})$ $WW$ fusion
amplitudes (the former computed with a 100 GeV Higgs boson).  Since
jet tagging was not assumed, $\overline qq \ra WZ$ is also included as
a background to the WZ fusion signal. Surprisingly, $\overline qq \ra
W^+Z$ is also a big background for $W^+W^+$ fusion, due to events in
which the negative lepton from the $Z$ escapes
detection.\cite{atlas} Similarly, it turns out that $\overline qq
\ra W^+\gamma^*$ is an equally important background.\cite{mcwk2} Top
quark backgrounds, $\overline tt$ and $\overline ttW$, have also been
considered and are easily controlled.
\begin{figure}
\begin{center}
{\epsfxsize= 0.9\textwidth \epsfbox {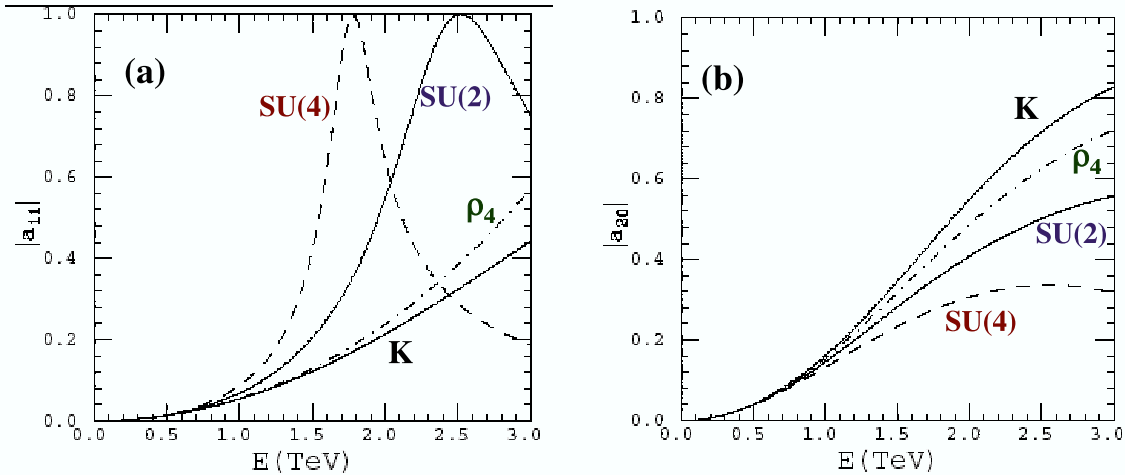}}
\end{center}
\caption{Partial wave amplitudes from technicolor and K-matrix 
models.\cite{mcwk1}}
\end{figure}

While a forward jet tag may eventually prove to be more
effective, the results quoted below from \cite{mcwk1,mcwk2} rely only on
hard lepton cuts and a central jet veto (CJV). The CJV vetos events
containing a jet with central rapidity, $\eta_{J}<2.5$, and high
transverse momentum, $P_{T}(J) > 60$ GeV; it reduces backgrounds from
transversely polarized $W$ bosons, which are emitted at larger
transverse momenta than the longitudinally polarized $W$ bosons of the
signal.  The hard lepton cuts rely on the general property that the
strong scattering cross sections are proportional to $s_{WW}$ while
the backgrounds scale like 1/$s_{WW}$, and on the differing
polarization of the signal and background $WW$ pairs.  If this
strategy suffices it has the advantage of being cleaner than relying
on forward jet tagging, which is subject to QCD corrections and to
detector-specific jet algorithms and acceptances in the forward
region. The results quoted below incorporate reasonable estimates of
the experimental efficiencies --- see \cite{mcwk1,mcwk2} for
details. The leptonic cuts are optimized separately for each set of
model parameters.

A robust observability criterion is defined and the cuts are optimized
by searching over the cut parameter space for the set of cuts that
satisfy the observability criterion with the smallest integrated
luminosity.  The criterion is
$$
\sigma^{\uparrow}   =  S/\sqrt{B}  \ge  5 
\eqno(6.1)
$$
$$
\sigma^{\downarrow}   =  S/\sqrt{S+B}  \ge  3
\eqno(6.2)
$$
$$
S \ge B,
\eqno(6.3)
$$
where $S$ and $B$ are the number of signal and background events, and
$\sigma^{\uparrow}$ and $\sigma^{\downarrow}$ are respectively the
number of standard deviations for the background to fluctuate up to
give a false signal or for the signal plus background to fluctuate
down to the 
\begin{figure}
\begin{center}
{\epsfxsize= 0.9\textwidth \epsfbox {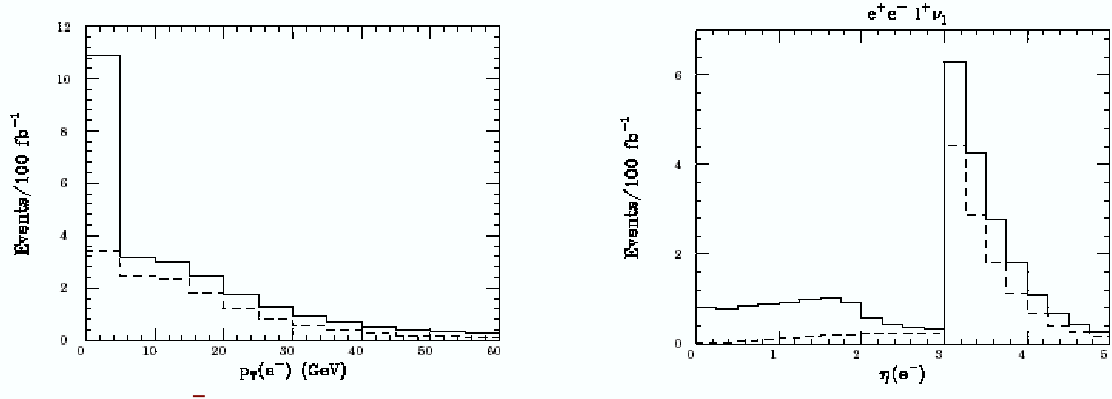}}
\end{center}
\caption{$p_T$ and $\eta$ distributions for wrong-sign leptons that 
escape veto.\cite{mcwk2}}
\end{figure}
level of the background alone.  The requirement $S \ge B$ is imposed so
that the signal is unambiguous despite the systematic uncertainty in
the size of the backgrounds; this condition might eventually be replaced 
by a less conservative one, since background studies {\it in situ} 
with real LHC data should substantially reduce the systematic 
uncertainties.

To obtain the $W^+W^+$ background the complete amplitude for
$\overline qq \rightarrow l^+ \nu l^+ l^-$ was computed in
\cite{mcwk2}, including amplitudes which are neither $\overline qq
\rightarrow WZ$ or $\overline qq \rightarrow W\gamma^*$. They are an
inevitable background because any detector has unavoidable blind spots
at low transverse momentum and at high rapidity where the $l^-$
escapes detection. At very low $p_T$, muons will not penetrate the
muon detector, electrons or muons may be lost in minimum bias pile-up,
and for low enough $p_T$ in a solenoidal detector they will curl up
unobservably within the beam pipe.  Muon and electron coverage is also
not likely to extend to the extreme forward, high rapidity region.

In reference \cite{mcwk2} an attempt was made to employ reasonable 
though aggressive assumptions about the observability of the extra 
electron or muon.  Rapidity coverage for electrons 
and muons was 
assumed for $\eta(l) < 3$.  Within this rapidity range it was assumed 
that isolated $e^-$ and $\mu^-$ leptons with $p_T(l) > 5$ GeV can be 
identified in events containing two isolated, 
central, high $p_T$ 
$e^+$'s and/or $\mu^+$'s.  It was also assumed that electrons (but not 
muons) with $1< p_T(l) < 5$ GeV can be identified if they are 
sufficiently collinear ($m(e^+e^-) < 1$ GeV) with a hard positron in 
the central region.  For $p_T(e^-) < 1$ GeV electrons were considered 
\begin{small}
\begin{quotation}
\noindent {\bf Table 6.1} Minimum integrated luminosity 
${\cal{L}}_{MIN}$ to satisfy significance criterion for $W^+W^+ + 
W^-W^-$ and $W^{\pm}Z$ scattering.
\end{quotation}
\end{small}
\begin{center}
\begin{tabular}{cccc}
$m_{\rho}$(TeV) & 1.8 &2.5&4.0\cr
\hline
\hline
${\cal{L}}_{MIN}(WW)\ ({\rm{fb}}^{-1})$&200&150&110\cr
\hline
${\cal{L}}_{MIN}(WZ)\ ({\rm{fb}}^{-1})$&44&320&NS\cr
\hline
\hline
\end{tabular}
\end{center}
to be unobservable. These assumptions should be reconsidered in view 
of the now finalized designs of ATLAS and CMS. Figure 7 shows the 
$p_T$ and $\eta$ distributions of the wrong-sign lepton for events 
in which it escapes the veto and the two like-sign leptons are in 
the signal region. 

The results are summarized in table 6.1, where it is assumed that a
high $p_T$ electron or muon can be detected with 85\% efficiency, a
high $p_T$ $Z$ decaying to electrons or muons with 95\% efficiency,
and that the $\overline qq \ra WZ/W\gamma^*$ background to the
like-sign $WW$ signal is rejected with 95\% efficiency when the
wrong-sign 

\noindent lepton falls within the acceptance region defined above.
(In \cite{mcwk2} veto efficiencies of 90 and 98\% were also
considered.) The 1.8 TeV $\rho$ meson of $N_{TC}=4$ technicolor
provides a signal in the $WZ$ channel that meets the observability
criterion with 44fb$^{-1}$. For the 4 TeV $\rho$ model, the resonance 
enhancement is unobservable and even for arbitrarily large luminosity 
there is no set of cuts that meet the observability criterion in 
the $WZ$ channel. But for this case the like sign $WW$ signal meets 
the criterion with 110 fb$^{-1}$. The worst case is intermediate 
between these two extremes: the $N_{TC}=2$ model with $m_{\rho} = 
2.5$ TeV is best detected in the like sign $WW$ channel for which 
150 fb$^{-1}$ is required. In this sense we may say that 
150 fb$^{-1}$ is the ``No-lose'' luminosity needed to assure the 
discovery of electroweak symmetry breaking in the worst case. 

For the $m_{\rho} = 2.5$ TeV worst case, the dominant background to
the $W^+W^+ +W^-W^-$ signal is from the $\overline qq \ra
WZ/W\gamma^*$ process: it accounts for 70\% of the background while
the $O(\alpha_W^2)$ and $O(\alpha_W\alpha_S)$ amplitudes contribute 26
and 3\% respectively. Forward jet tagging would eliminate the
$\overline qq \ra WZ/W\gamma^*$ background, so a lower ``No-lose''
luminosity might well be attainable. The $WZ$ signal might also be
improved if it turns out that the mixed modes are practicable, in
which the $W$ decays hadronically while the $Z$ decays leptonically.

Theorist-level estimates are of course oversimplified and optimistic.
For instance, apart from guesses at the efficiencies there is no
attempt to simulate the detector, to
model the 
effect of pileup or charge misidentification....
Nevertheless it is likely after the LHC begins to run that
experimenters will devise aggressive methods that will ultimately
yield even better results: they will be free to try  
bold strategies because they will be able to test them 
with real data. An example from the past is provided by the
experience at LEP I, where experimenters rapidly exceeded the reach
for the Higgs boson that was projected in the first CERN yellow book
study for LEP, because they were able to show with real data
that $Z^* \ra H \overline{\nu}\nu$ is a viable detection mode, which
must have seemed too adventurous to the yellow book authors.

\noindent {\bf 7. The bottom line}

The origin of electroweak symmetry breaking is completely unknown ---
many possibilities are open, among them undoubtedly some not yet 
imagined. With enough luminosity and experimental ingenuity, the 
LHC at design luminosity and energy is sure to lead us to the answer. 
To cover all possibilities it is essential to develop the capability 
to observe strong $WW$ scattering if it exists or to exclude it if it 
does not.


{\bf Acknowledgements} I wish to thank Gustavo Burdman, Yasunori
Nomura, Ian Hinchliffe, John Terning, Georges Azuelos, Pawel Zych, and
Jonathan Bagger for helpful discussions.


{\small This work was supported in part by the Director, Office of
Science, Office of High Energy and Nuclear Physics, Division of High
Energy Physics, of the U.S. Department of Energy under Contract
DE-AC03-76SF00098}


\end{document}